\documentclass[twocolumn,secnumarabic,amssymb, nobibnotes]{revtex4}
\usepackage[T1]{fontenc}
\usepackage{color}
\usepackage{dcolumn}
\usepackage{textcomp}
\usepackage{amsmath}
\usepackage{graphicx}
\usepackage{epstopdf}
\usepackage{textcomp}
\usepackage{subfigure}
\usepackage{hyperref}
\usepackage{caption}
\usepackage{enumitem}
\usepackage{mathrsfs}
\usepackage{epstopdf}
\usepackage{graphicx}
\usepackage{revsymb}
\usepackage{subfigure}
\usepackage{amsbsy}
\usepackage{amsfonts}
\usepackage{amsthm}
\usepackage{float}
\usepackage{natbib}
\usepackage{epstopdf}
\usepackage[normalem]{ulem}

\newtheorem{theorem}{Theorem}

\newtheorem{remark}{Remark}[theorem]

\begin{document}

\title{Quantum no-go theorems in causality respecting systems in presence of\\closed timelike curves: Tweaking the Deutsch condition}
\author{Asutosh Kumar\(^{1,2}\),  Indranil Chakrabarty\(^3\), Arun Kumar Pati\(^{1}\), Aditi Sen(De)\(^{1}\), Ujjwal Sen\(^{1}\)}
\affiliation{\(^1\)Harish-Chandra Research Institute, HBNI, Chhatnag Road, Jhunsi, Allahabad 211 019, India}
\affiliation{\(^2\)P.G. Department of Physics, Gaya College, Magadh University, Rampur, Gaya 823 001, India}
\affiliation{\(^3\)Center for Security, Theory and Algorithmic Research, International Institute of Information Technology-Hyderabad, Gachibowli, Hyderabad, India.}

\begin{abstract} 

\noindent We consider causality respecting (CR) quantum systems interacting with closed timelike curves (CTCs), within the Deutsch model. 
We introduce the concepts of popping up and elimination of quantum information and use them to show that no-cloning and no-deleting, which are 
true in CR quantum systems, are no more valid in the same that are interacting with CTCs. We also find limits on the possibility of 
creation of entanglement between a CR system and a CTC, and the same between two CR systems in the presence of a CTC. We prove 
that teleportation of quantum information, even in its approximate version, from a CR region to a CTC is disallowed.
Interestingly, we find that tweaking the Deutsch model, by allowing the input and output to be not the same, leads to a nontrivial approximate teleportation beyond the classical limit.

\end{abstract}
\maketitle 

\section{Introduction}

\noindent No-go theorems play an important role in quantum information science \cite{Nielsen-Chuang}. 
They are crucial both in nonclassical applications of quantum states and operations as resources, and 
towards a better understanding of quantum concepts. 
A good example is the no-cloning theorem \cite{wz, dieks, yuen}, which states that nonorthogonal quantum states cannot be cloned perfectly, and can be seen as an underlying feature of the security of quantum cryptography \cite{Gisin-rmp-crypto}. 
Other examples of no-go theorems like no-deleting, no-broadcasting, no-splitting, no-hiding, etc. can be found in Refs. \cite{broadcasting, superbroadcasting, pb, ref-no-split, no-partial-erasure, no-hiding}.\\


\noindent Recently, there have been some exciting developments in quantum information 
theory in the presence of closed timelike curves (CTCs). It is known that the 
general theory of relativity does allow the existence of a CTC, 
a world line which connects back on itself \cite{stockum, godel, deser, ori, mor,fro,kim,got,haw, politzer} (see \cite{mor, haw} however).
An important question is whether  one can formulate a consistent 
theory of quantum mechanics in the presence of CTCs. Such a formalism was 
developed by Deutsch who proposed a  
model of quantum theory in the presence of CTCs \cite{deu} (cf. \cite{pctc1, pctc2, pctc3, pctc4, pctc5, pctc6, pctc7}).\\

\noindent 
Investigations have been made to see how the presence of closed timelike curves can affect the computational 
power
and other abilities to perform information processing tasks of a system 
\cite{bru,bac,aar}. 
From the perspective of quantum computation, 
it has, e.g., been shown that  access to a CTC would allow factorization of composite numbers
efficiently with the help of a classical computer \cite{bru}, 
a CTC-assisted
quantum computer would be able to solve NP-complete problems \cite{bac}, both classical and quantum computers under CTC belong to the same complexity class \cite{aar}, etc.
%
On the other hand, Brun {\em et al.} \cite{bru1} have shown that if one 
has access to CTCs, then one can perfectly
distinguish nonorthogonal quantum states, having possible implications for the security of quantum cryptography \cite{Gisin-rmp-crypto}. 
In the presence of CTCs,  
information flow of quantum states \cite{ralph}, 
purifications of mixed states \cite{akp}, and  nonlocal boxes \cite{indra} have also been addressed.
Bennett {\it et al.} \cite{ben1} have argued that the implications obtained by assuming the existence of CTCs need to be revisited when we assume that the input state 
in any protocol is a mixture 
of the possible inputs.\\

\noindent Before presenting the results, let us briefly describe here the Deutsch formalism \cite{deu}. It involves a
unitary interaction, \(U\), between a quantum system in a causality-respecting (CR) region
with another system  that has  a world line in a region 
where closed timelike curves exist. It is assumed that the states of these systems  
are  density matrices in standard quantum
mechanics. The combined state of the CR and CTC systems before the interaction is a product state, $\rho_{\rm CR}\otimes \rho_{\rm CTC}$.
The unitary transformation on the joint system
changes the composite state as
\begin{equation}
\label{eq-dadu}
\rho_{\rm CR}\otimes \rho_{\rm CTC} \rightarrow 
U (\rho_{\rm CR}\otimes \rho_{\rm CTC}) U^{\dagger}. 
\end{equation}
For self-consistency, the Deutsch model requires
\begin{equation}
\label{eq-nati}
{\rm  Tr}_{\rm CR}(U \rho_{\rm CR}\otimes \rho_{\rm CTC} U^{\dagger})=\rho_{\rm CTC},
\end{equation}
the ``Deutsch self-consistency condition''. 
Therefore the Deutsch formalism demands that the state of the CTC system at the output of  the evolution is the same as 
that of the CTC system at the input of the evolution. Mathematically, the state of the CTC system is a fixed point solution of the self-consistency equation.
In case there are multiple fixed points, one chooses their
maximum-entropy mixture. On the other hand, the final state of the 
CR system, which does not have such restriction imposed on it, is given by
\begin{equation}
\label{eq-hati}
{\rm Tr}_{\rm CTC}(U \rho_{\rm CR}\otimes \rho_{\rm CTC} U^{\dagger}) =
\rho{'}_{\rm CR}.
\end{equation}
The equations (\ref{eq-dadu}), (\ref{eq-nati}), and (\ref{eq-hati}) form the basic equations that  provide the entire dynamics 
of the composite CTC and CR systems.
It is important to note that $\rho'_{\rm CR}$ depends nonlinearly 
on $\rho_{\rm CR}$. Thus, it is evident that the output of the CR system is a nonlinear 
function of the input CR system density matrix \cite{cas}.\\


 \noindent 
In this work,  we uncover  two exotic features namely the ``popping up'' and ``elimination'' of quantum information in the  presence of CTCs. We show that the information present in the CR system ``pops up'' in the CTC qubit 
when it is allowed to interact with the CR system. We  also show that it is possible to destroy or completely ``eliminate'' the information of the state of the CR system in presence of CTCs. 
Using these twin features, we prove  the violations of two fundamental theorems of quantum mechanics, 
namely, the no-cloning \cite{wz, dieks, yuen, bh1, bh2, bh3, bh4, bh5, bh6} and the no-deleting \cite{pb, sa1, sa2} theorems. Note that  cloning of an arbitrary quantum state of a CR system when in contact with a CTC is already known \cite{deu, plas, ahn, brun}.
We provide an independent proof of the same, which, we believe, gives a fresh perspective on the mechanism of the phenomenon of cloning.
While cloning and deleting of quantum systems become possible in CR systems when interactions with CTCs are allowed, teleportation \cite{tele}, 
a phenomenon that plays an important 
role in several aspects in the quantum mechanics of CR systems, becomes impossible. We first show the limits within which entanglement can be 
generated between a CR system and a CTC, and subsequently prove that teleportation is always disallowed. In contrast to, e.g., the no-cloning theorem in 
quantum theory of CR systems, the ``no teleportation theorem'' here does not allow even an approximate teleportation. 
We however show 
that  ``tweaking''
 the Deutsch criterion that  allows  
deviations of the final states from their corresponding initial states, 
can result in 
nonclassical teleportation fidelities. 

Our paper is organized as follows.
The concepts and the corresponding proofs about popping up  and elimination are presented in Sec. \ref{sec-pop-eli}.
The possibility of cloning and deleting using CTCs are presented in Sec. \ref{sec-clo-del}.
The results 
on entanglement and teleportation are presented in Sec. \ref{sec-ent-tel}. The concept of  ``tweaking''
 the Deutsch criterion, and the possibility of nontrivial teleportation obtained thereby are presented in  Sec. \ref{sec-tweak}.
We present a  conclusion in Sec. \ref{sec-conclu}.

\section{Popping Up and Elimination of Quantum Information}
\label{sec-pop-eli}
\noindent There exists a notion of ``permanence'' of quantum information in the sense that information in a quantum world can neither be created nor destroyed \cite{Jozsa}. In this section, we show that in the presence of a closed timelike curve, the information present in a CR system 
``pops up'' in the CTC, and on the other hand, 
the information of a state present in a CR system can be completely destroyed (``eliminated'').
These results suggest that quantum information in the presence of CTCs does not respect permanence of information.\\

\noindent
{\bf Popping Up of Quantum Information:}\\


\noindent Let $|\psi \rangle_{\rm CR} \langle \psi|$ be an arbitrary state of the 
CR system 
 and 
$\rho_{\rm CTC}$ 
be the state of the CTC. Let us suppose that they interact via the unitary, $U_{\rm SWAP}$,  
the swap operation,
defined, for the bi-orthonormal product basis, \(\{|i\rangle_{\rm CR} |j\rangle_{\rm CTC}\}_{i,j}\), as 
\(U_{\rm SWAP} |i\rangle_{\rm CR} |j\rangle_{\rm CTC} = |j\rangle_{\rm CR} |i\rangle_{\rm CTC}\). 
We will often denote $|\psi \rangle \langle \psi|$ as $\psi$. See Fig. \ref{fig-popping-elimination}(a) for circuit diagram of popping up of quantum information.
Hence, we have 
%
\begin{eqnarray}
\psi_{\rm CR} \otimes \rho_{\rm CTC} 
\overset{U_{SWAP}} \rightarrow
\rho_{\rm CR} \otimes \psi_{\rm CTC}.
\label{Eq-RumiDarwaza}
\end{eqnarray}
Now, if we trace out the CR system, the final state of the CTC is
$|\psi \rangle_{\rm CTC} \langle \psi|$. By the 
consistency 
condition of 
Deutsch, this should be  same as the original state of the CTC, 
i.e., $\rho_{\rm CTC} = |\psi \rangle_{\rm CTC} \langle \psi|$.
This equation is now a mathematical relation on the space of operators on the CTC Hilbert space. As a mathematical relation, it remains valid in the CR system also.
We therefore have that the joint CR-CTC system evolves as 
\begin{eqnarray}
\psi_{\rm CR} \otimes \rho_{\rm CTC} \rightarrow 
\psi_{\rm CR} \otimes \psi_{\rm CTC},
\end{eqnarray}
implying that  the information in the CR system is ``popping up'' in the CTC system. 
What is more, considering $\rho_{\rm CTC} = |\psi \rangle_{\rm CTC} \langle \psi|$ to be a valid mathematical relation after the swap operation has been applied,  
we 
can also say that to start with we had the CR and the CTC systems in the state 
$|\psi \rangle \langle \psi| \otimes |\psi \rangle \langle \psi|$. Since this 
holds for an arbitrary input state of the CR system, the CTC
has knowledge about all possible states in the CR system. Thus the mere possibility that a CTC  \emph{will be interacted} with a CR system can result in the popping up 
of the quantum information of the CR system in the CTC. 
Note that the popping up  happens even before the interaction is implemented.
Hence a 
CTC has knowledge about the past and future of all CR quantum systems in the universe, provided we assume that there exists a CTC that can be interacted with 
the relevant CR system with the swap operation.

\begin{remark}
Note that the conclusion that $\rho$ and $\psi$ are equal is a result of the Deutsch consistency equation, and our choice of SWAP as the interaction between the CR system and CTC.  Also, the above discussion of ours is for the situation when we choose SWAP as our interaction unitary. The discussion is not valid in general. If another unitary is chosen, the conclusion could be different.
\end{remark}

\noindent 
While the discussions in the preceding paragraph were when the CR system is in a pure state, the same line of reasoning goes through if the CR system is in a mixed 
state, which in turn can be part of an entangled state \cite{HHHH-rmp}.
\\

\noindent \textbf{Elimination of Quantum Information:}\\

\noindent Just like creation of quantum information, CTCs can assist in the complete destruction of 
quantum information in a CR system. 
However, unlike in popping up, here we require two CR systems $|\psi \rangle_{\rm CR(1)} \otimes 
|0\rangle_{\rm CR(2)}$ and a CTC system $\rho_{\rm CTC}$, which interact with the unitary $U_{\rm SWAP}$.  Note that $|0\rangle$ is a fixed state. We denote $|\psi \rangle_{\rm CR} \langle \psi|$ as $\psi_{\rm CR}$, and so on. See Fig. \ref{fig-popping-elimination}(b) for circuit diagram of elimination of quantum information.
Applying $U_{SWAP}$ between CR(2) and CTC, we have
\begin{eqnarray}
\label{eq-elim1}
\psi_{\rm CR(1)} \otimes 0_{\rm CR(2)} \otimes \rho_{\rm CTC}  \rightarrow 
\psi_{\rm CR(1)} \otimes \rho_{\rm CR(2)} \otimes 0_{\rm CTC}.  
\end{eqnarray}
Applying the Deutsch self-consistency condition, we get $\rho = 0$.
Therefore, after the action of the swap operation and the application of the Deutsch condition, we have
\begin{eqnarray}
\label{eq-elim2}
\psi_{\rm CR(1)} \otimes 0_{\rm CR(2)} \otimes \rho_{\rm CTC}  \rightarrow 
\psi_{\rm CR(1)} \otimes 0_{\rm CR(2)} \otimes 0_{\rm CTC}.  
\end{eqnarray}
We now apply $U_{SWAP}$ between CR(1) and CTC to obtain
\begin{eqnarray}
\label{eq-elim3}
\psi_{\rm CR(1)} \otimes 0_{\rm CR(2)} \otimes 0_{\rm CTC} \rightarrow 
0_{\rm CR(1)} \otimes 0_{\rm CR(2)} \otimes \psi_{\rm CTC}.
\end{eqnarray}
If we are allowed to apply the Deutsch condition once more (see \cite{imp-note} for our viewpoint), we obtain $0 =\psi$, so that the final state of the three-party system is 
\begin{eqnarray}
\label{eq-elim4}
0_{\rm CR(1)} \otimes 0_{\rm CR(2)} \otimes 0_{\rm CTC}, 
\end{eqnarray}
so that the information about the input state $\psi$ in CR(1) has been completely eliminated. Similar to the case of popping up, the CR system can also be initially in a mixed state. 
\begin{widetext}
\begin{center}
\begin{figure}[htb]
\includegraphics[scale=0.3]{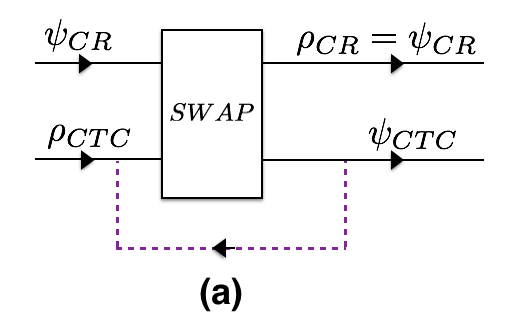}
\includegraphics[scale=0.3]{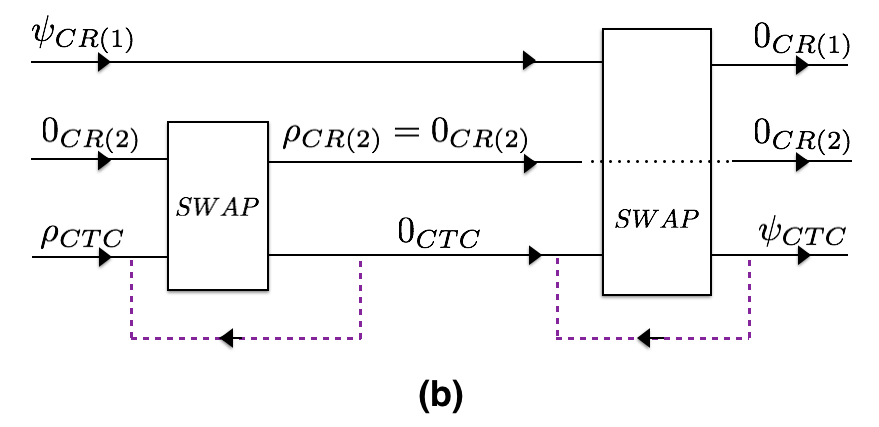}
\caption{Circuit diagrams of (a) popping up and (b) elimination, of quantum information. The broken lines with arrow in opposite direction depicts the dynamics of a CTC, satisfying the Deutsch self-consistency condition Eq.~(\ref{eq-nati}). Here $SWAP$ is the required unitary operation.}
\label{fig-popping-elimination}
\end{figure}
\end{center}
\end{widetext}

\section{Cloning and Deleting of arbitrary quantum state in presence of CTCs}
\label{sec-clo-del}
\noindent Cloning and deleting of arbitrary quantum states are not possible in the quantum world \cite{wz, dieks, yuen, bh1, bh2, bh3, bh4, bh5, bh6, pb, sa1, sa2}. 
In this section, we use the phenomena of popping up and elimination of quantum information to prove that 
 it is always possible to clone and delete quantum states in presence of CTCs. The possibility of cloning of quantum states  in the presence of a CTC 
is known \cite{deu, plas, ahn, brun}, and Ref. \cite{ahn}, in particular,  uses a sequence of unitaries including the swap operation to achieve the same. 
Here we present an independent proof that 
reveals  an interesting way in which cloning can take place in a world with CTCs.
It is to be noted here that the phenomena of popping up  or elimination by themselves are not cloning or deleting (respectively), as we want the entire phenomena of cloning or deleting to occur only in CR systems. 
\\

\noindent {\bf Cloning of an Arbitrary State:} \\

\begin{theorem}
\label{thm-cloning}
If $|\psi \rangle_{\rm CR(1)} $ is an arbitrary quantum state in a CR region, then 
there exists a map $|\psi \rangle_{\rm CR(1)}  \rightarrow |\psi \rangle_{\rm CR(1)} |\psi \rangle_{\rm CR(2)} $ from one CR region to another, in the presence of a CTC
satisfying the Deutsch condition for the swap operator.
\end{theorem}

\begin{proof}
Consider two CR systems, one in an arbitrary quantum state,
$|\psi \rangle_{\rm CR(1)}$, 
(the state to be cloned), and another in a fixed ``blank'' quantum state, $|B \rangle_{\rm CR(2)} $.
Consider also a CTC in the state \(\rho_{\rm CTC}\). The CTC system can be thought of as the ``machine''.
We will apply two swap gates to prove the possibility of cloning in this model. Let us begin with the first, which is applied on 
the systems CR(1) and CTC, whence 
\begin{eqnarray}
&\psi_{\rm CR(1)} \otimes B_{\rm CR(2)} \otimes \rho_{\rm CTC}  \nonumber \\
&\overset{U_{SWAP}^{CR(1):CTC}} \rightarrow
\rho_{\rm CR(1)} \otimes B_{\rm CR(2)} \otimes \psi_{\rm CTC}.
\end{eqnarray}
Applying the Deutsch  condition, we have 
$\rho = \psi$ as a mathematical relation which can be applied to the CR system, so that the state of the tripartite system becomes 
$\psi_{\rm CR(1)} \otimes B_{\rm CR(2)} \otimes \psi_{\rm CTC}$.
This is the phenomenon of popping up between the systems CR(1) and CTC.
We now apply the second swap operation on the systems CR(2) and CTC, whereby
\begin{eqnarray}
\label{eq-11}
&\psi_{\rm CR(1)} \otimes B_{\rm CR(2)} \otimes \psi_{\rm CTC}  \nonumber \\
&\overset{U_{SWAP}^{CR(2):CTC}} \rightarrow
\psi_{\rm CR(1)} \otimes \psi_{\rm CR(2)} \otimes B_{\rm CTC}.
\end{eqnarray}
\end{proof}


\begin{remark}
After the first swap operation in the proposed cloning process, the CTC is in a fixed state $\psi$ (which is the initial state of the system whose state is to be cloned). At this point, there are at least two options to take. We may assume that when we apply the swap operation for the second time, we do not apply the Deutsch consistency condition. In that case, the right-hand side of Eq. (\ref{eq-11}) will be $\psi \otimes \psi \otimes B$, and this is what it is now. 
However, we can choose a second option. We can assume that the Deutsch condition has to be applied every time we consider a dynamics including a CTC. In that case, applying the Deutsch condition after we act with the swap operator for the second time, we obtain $\psi = B$. Thereby, the right-hand side of Eq. (\ref{eq-11}) could be any one of the eight possible combinations from the set $\{\psi \otimes \psi \otimes \psi, \psi \otimes \psi \otimes B, \ldots, B \otimes B \otimes B\}$. We choose $\psi \otimes \psi \otimes B$, as this achieves our goal of cloning.
\end{remark}

\noindent{\bf Deleting of an Arbitrary State:}\\

\noindent In the case of deleting, 
we are given two copies of an arbitrary state \(|\psi\rangle\), and we ask if there can be a
physical (quantum) operation in a closed system which can take  
$|\psi \rangle | \psi \rangle |A \rangle \rightarrow 
|\psi \rangle | 0 \rangle |A{'} \rangle$, where \(| 0 \rangle\), \(|A \rangle\), and \(|A{'} \rangle\) are fixed states independent of 
\(|\psi\rangle\). The no-deleting theorem says that this operation is impossible in the quantum world, where the third subsystem consists of the entire universe 
outside the first two subsystems.
%
However, it follows from the discussion above on elimination of quantum information that in the presence of CTC, deleting is possible simply by elimination of quantum information in the second copy of $\psi$, while the first copy stands idle. More precisely, we have the following result
 
\begin{theorem}
\label{thm-del}
In a closed CR quantum system, deleting is possible provided it is allowed to interact with a CTC system satisfying the Deutsch condition for the swap operator.
\end{theorem}

\begin{proof}
We begin with two copies of $|\psi \rangle$ and a single copy of $|0\rangle$ in CR regions, and CTC in the state 
 $\rho$. 
We denote $|\psi \rangle_{\rm CR} \langle \psi|$ as $\psi_{\rm CR}$, and so on. We again assume that the Deutsch self-consistency condition can be called more than once.
The proof can be gathered in following lines: 
\begin{eqnarray}
\label{eq-del}
&\psi_{\rm CR(1)} \otimes \psi_{\rm CR(2)} \otimes 0_{CR(3)} \otimes \rho_{\rm CTC} \nonumber \\
& \overset{U_{SWAP}^{CR(3):CTC}} \longrightarrow 
\psi_{CR(1)} \otimes \psi_{\rm CR(2)} \otimes \rho_{CR(3)} \otimes 0_{CTC} \nonumber \\
& \overset{(\rho = 0)} = \psi_{CR(1)} \otimes \psi_{\rm CR(2)} \otimes 0_{CR(3)} \otimes 0_{CTC} \nonumber \\
& \overset{U_{SWAP}^{CR(2):CTC}} \longrightarrow 
\psi_{\rm CR(1)} \otimes 0_{CR(2)} \otimes 0_{CR(3)} \otimes \psi_{CTC} \nonumber \\
& \overset{(\psi = 0)} = \psi_{\rm CR(1)} \otimes 0_{CR(2)} \otimes 0_{CR(3)} \otimes 0_{CTC},  
\end{eqnarray}
where the equalities follow due to the Deutsch self-consistency condition.
Therefore, we have been able to take the CR(2) system to a standard state, independent of its input state \(|\psi\rangle\), while still keeping the state of the CR(1) system at its pre-interaction state. Furthermore,
 the CTC system is finally in the state $|0\rangle$, which is independent of $|\psi\rangle$, so that there is no information that is leaked out from the CR systems, CR(1) and CR(2), into the CTC. 
\end{proof}

\begin{remark}
More specifically, the discussion on elimination of quantum information implies that if $|\psi \rangle $ and \(|0\rangle\) are respectively arbitrary and fixed quantum states, then there exists a map 
$|\psi \rangle_{\rm CR(1)} |\psi \rangle_{\rm CR(2)} \rightarrow |\psi \rangle_{\rm CR(1)} |0 \rangle_{\rm CR(2)} $ of two CR regions,
in presence of a CTC
satisfying the Deutsch condition for the swap operator and an additional CR system initially in the state $|0\rangle$. The action can be considered to be closed as the final state of the CTC system, the additional CR system,  or any other system does not retain any information of the initial states of the CR systems, CR(1) and CR(2).
This is important as deleting must be considered in a system from which information is not leaking into the environment \cite{pb,sen,sen1}. 
\end{remark}

\begin{remark}
We note here that both in the case of cloning and deleting, we require to apply the swap operator twice. However, while we call the Deutsch self-consistency condition once for cloning, it has to be used twice for deleting. 
\end{remark}

\section{Impossibility of Teleporting a Quantum State from CR to CTC} 
\label{sec-ent-tel}

\noindent \textbf{Creating Entanglement between CR and CTC}:\\

\noindent Before considering the question of teleportation between CR and CTC worlds, let us briefly consider whether  entanglement can be created between these two worlds, a question addressed before in Ref. \cite{akp}. Suppose that we wish to create the state \(\rho_{\rm CR(1),CTC}\), 
which may be entangled, between a CR system, CR(1), and a CTC. 
For this purpose, we consider two CR systems and a CTC to be initially in the state \(\rho_{\rm CR(1),CR(2)} \otimes \tilde{\rho}_{\rm CTC}\).
Applying the swap operator, \(U_{\rm SWAP}^{\rm CR(2):CTC}\), we will be able to have the state \(\rho_{\rm CR(1),CTC}\), irrespective of whether 
the bipartite state \(\rho\) is entangled or separable. Note that for the Deutsch condition to be satisfied for the evolution considered, one must have $\tilde{\rho}_{\rm CTC} = {\rm Tr}_{\rm CR(1)} \rho_{\rm CR(1),CR(2)}$.\\
In spite of this possibility, we have the following no-go theorems. 
\\

\begin{theorem}
\label{thm-no-entanglement1}
{\bf (No Entanglement between CR and CTC)} A CR system and a CTC cannot get correlated, neither classically nor quantumly, if the CTC is allowed to interact with another CR system in a fixed pure state, and if the CTC follows the Deutsch condition for the swap operator.
\end{theorem}

\begin{proof}
Suppose that we are given a CR system and a CTC in the state \(\rho_{\rm CR(1),CTC}\), and another CR system in the pure state \(|0\rangle_{\rm CR(2)} \langle 0|\) is allowed 
to interact with CTC by the swap operator. Post-interaction, we apply the Deutsch condition. This implies that the local state of CTC must be (both before and after the 
interaction) in the \emph{pure} state \(|0\rangle\). Thus,  \(\rho_{\rm CR(1),CTC}\) must have already been in a product state, precluding any correlation between the CR system, CR(1), and the CTC. 
\end{proof}

\begin{theorem}
\label{thm-no-entanglement2}
{\bf (No Entanglement between two CRs)} Two CR systems, CR(1) and CR(2), cannot get correlated, neither classically nor quantumly, if (a) a third CR system in a fixed pure state is allowed to interact with a CTC via the swap operator, and if (b) after the operation in (a), either CR(1) or CR(2) is allowed to interact with the CTC via the swap operator, and if we are allowed to apply the Deutsch condition for the second swap operation. 
\end{theorem}

\begin{proof}
Suppose that initially two CR systems are in the state \(\rho_{\rm CR(1),CR(2)}\), a CTC system is in the state \(\tilde{\rho}_{\rm CTC}\), and a third CR system is in the state \(|0\rangle_{\rm CR(3)} \langle 0|\).  Now the CTC system is allowed to interact via the swap operator with CR(3), followed by with CR(2). Post-interaction, if we apply the Deutsch condition, we get \(|0\rangle_{\rm CTC} \langle 0| = {\rm Tr}_{\rm CR(1)} \rho_{\rm CR(1),CTC}\). This implies that there is no correlation between two CR systems, CR(1) and CR(2).
\end{proof}

\begin{remark}
While we refer to the above theorems (Theorems \ref{thm-no-entanglement1} \& \ref{thm-no-entanglement2}) as the ``no entanglement theorems'', they are in reality ``no correlation theorems'', as neither classical nor quantum correlations are allowed in the situations described. We call them as the ``no entanglement theorems'', since we are using them in the context of the no teleportation theorem below. 
%
We would also like to mention that popping up of information in a CR-CTC system, if it were possible in a situation where the dynamics is linear, would create classical correlations between the CR system and the CTC. Because, in such a (hypothetical) case, one can use a mixture of two states, $|\psi\rangle$ and $|\phi\rangle$ (say), as the input for the CR system, and the output of the joint CR-CTC system will then be in a classically correlated state (mixture of $|\psi\rangle |\psi\rangle$ and $|\phi\rangle |\phi\rangle$). Note that we have assumed that the two copies of $|\phi\rangle$ (or $|\psi\rangle$) are created by the nonlinear evolution via the Deutsch condition, while the mixing is done linearly.
The evolution here is however not linear, and hence the proof of the implication is not valid. By this argument, we do not, however, claim that non-linearity implies no correlations. 
\end{remark}

\begin{remark}
We have discussed a method of creating entanglement between a CR system and a CTC, above Theorem \ref{thm-no-entanglement1}. Theorem \ref{thm-no-entanglement1} states that entanglement between a CR system and a CTC can exist only if the premises of this  theorem are not satisfied. Similarly, Theorem \ref{thm-no-entanglement2} states that the entanglement between the two CR systems can exist only if the premises of the theorem are not satisfied. 
\end{remark}

\noindent\textbf{No Teleportation of Quantum States from CR to CTC}:\\

\noindent Assuming that creation of an entangled state between a CR system and a CTC is possible, 
a next question is whether the same can be put to use. In particular, can we use the entangled state to teleport quantum information from 
the CR system to the CTC. The following result on teleportation is actually independent of whether the shared state is entangled or not.\\

\begin{theorem}
\label{thm-no-teleportation}
{\bf (No Teleportation Theorem)} It is not possible to teleport, even approximately, an arbitrary quantum state from a CR system to a CTC, if after the teleportation, the CTC is allowed to interact with another CR system in a fixed state via the swap operator, and if the Deutsch condition can be applied for the swap operator.
\end{theorem}

\begin{proof}
Consider first the case of exact teleportation. 
Suppose Alice and Bob share the state \(\rho_{\rm CR(2),CTC}\), of a CR system and a CTC. Let us suppose that Alice possesses the state \(\tilde{\rho}_{\rm CR(1)}\) to be teleported. In addition, assume that Bob has another CR system, in a fixed state \(\rho'_{\rm CR(3)}\), in his possession.  
Irrespective of whether the shared state between Alice and Bob is entangled or 
separable, let us assume that it is possible to exactly teleport the arbitrary state of the CR(1) system to the CTC, by using the shared state. That is,
\begin{eqnarray}
&\tilde{\rho}_{\rm CR(1)} \otimes \rho_{\rm CR(2),CTC} \otimes \rho'_{\rm CR(3)}  \nonumber \\
&\overset{Teleportation} \longrightarrow
\tilde{\tilde{\rho}}_{\rm CR(1),CR(2),CTC} \otimes \rho'_{\rm CR(3)}.
\end{eqnarray}
If we apply the Deutsch condition after the teleportation has been performed, we obtain
\begin{equation}
{\rm Tr}_{\rm CR(2)}\rho_{\rm CR(2),CTC} = {\rm Tr}_{\rm CR(1),CR(2)}\tilde{\tilde{\rho}}_{\rm CR(1),CR(2),CTC}. 
\end{equation}
However, we will not be using this equation. Though exact teleportation seems feasible here, it requires that post-teleportation, CTC part of the teleported state must be kept isolated, i.e., it must be prohibited to interact with any CR system in a fixed state because it would then render teleportation impossible, even approximately, as we show now.  
Let us assume that after teleportation, the CTC is allowed to interact with the CR(3) system via the swap operator, \(U_{\rm SWAP}^{\rm CR(3):CTC}\), which yields
\begin{equation}
\tilde{\tilde{\rho}}_{\rm CR(1),CR(2),CR(3)} \otimes \rho'_{\rm CTC}.
\end{equation}
If we apply Deutsch condition here, we get
\begin{equation}
{\rm Tr}_{\rm CR(1),CR(2)}\tilde{\tilde{\rho}}_{\rm CR(1),CR(2),CTC} = \rho'_{\rm CTC}.
\end{equation}
So the state obtained after teleportation is a fixed state. But the success of the teleportation protocol, in the exact case, implies that this fixed state must be equal to the \emph{arbitrary} state of CR(1). Therefore, exact teleportation is not possible.
Moreover, since the output CTC state, for the teleportation protocol, must be in a fixed state, irrespective of the input at CR(1), we cannot even have an approximate teleportation in the scenario considered. 

\end{proof}



\section{Tweaking the Deutsch condition and approximate teleportation}
\label{sec-tweak}

\noindent The no teleportation theorem, derived in the preceding section, 
is similar to the no-go theorems of quantum mechanics in causality respecting systems. 
However, it 
has a distinct character of its own. In particular, both exact and approximate teleportations are not allowed from CR systems to CTC. This is in sharp contrast 
to, e.g., the quantum no cloning theorem in CR systems, which disallows exact cloning but allows approximate cloning \cite{bh1, bh2, bh3, bh4, bh5, bh6}. In a quest for attaining approximate 
teleportation, here we address the question 
whether the same can be achieved by some change in the Deutsch condition (cf. \cite{bru1}). Through this exercise, we may also gain insight about physical phenomena when the Deutsch condition is not stringent. The immediate question, however, is whether a non-stringent Deutsch condition allows us 
to attain approximate teleportation of an arbitrary state \(|\psi\rangle\) in a CR system to a CTC, whereby an approximate copy, \(\rho_{\rm CTC}^f(\psi)\),
is created in the CTC, which is as close as possible to the arbitrary state \(|\psi\rangle\). 
We demand that the fidelity of the approximate copy (averaged over the input state space), defined as 
\begin{equation}
\mathcal{F} = \int \langle \psi | \rho_{\rm CTC}^f(\psi) |\psi \rangle d\psi,
\end{equation}
be higher than the classically achievable fidelity, i.e., the fidelity that is possible to achieve without the use of entangled shared states \cite{HHH-tele}. 
For the case of teleportation of a qubit, the  classically achievable fidelity is \(\frac23\) in the quantum domain.\\

\noindent Let us now introduce a deviation, which is just sufficient for the intended purpose of attaining nonclassical teleportation, 
of the Deutsch self-consistency condition. To do that, we suppose that the initial and final states of the CTC, respectively denoted by 
\(\rho_{\rm CTC}^i\) and  \(\rho_{\rm CTC}^f\), are such that 
\begin{equation}
 \| \rho_{\rm CTC}^f - \rho_{\rm CTC}^i \| \leq \epsilon,
\end{equation}
for an \(\epsilon\) in \([0,1]\). The norm is chosen in such a way that both 
\(\rho_{\rm CTC}^f - (1-\epsilon) \rho_{\rm CTC}^i\) and \((1+\epsilon) \rho_{\rm CTC}^i - \rho_{\rm CTC}^f\)
are positive semidefinite (see \cite{Simmons, wilde}). 
We refer to this as the \emph{\(\epsilon\)-close Deutsch model of CTC}, or, for short, {\em \(\epsilon\)-CTC}.\\

\noindent For simplicity, let us suppose that the final state, $\rho_{\rm CTC}^f$, of the CTC, after the teleportation protocol has been carried out,
 is of the form
\begin{equation}
\rho_{\rm CTC}^f=(1-\epsilon)\rho_{\rm CTC}^i+\epsilon \rho, 
\label{eq:final-state}
\end{equation}   
where $\rho$ is the arbitrary state to be teleported. 
For $\epsilon = 0$, the state is not teleported at all, and  the original Deutsch self-consistency condition is 
exactly satisfied, while when
$\epsilon = 1$, the state is perfectly teleported. Note that in all cases where \(\epsilon\) is nonzero, the Deutsch self-consistency is violated. 
Now while 
$\rho_{\rm CTC}^f-(1-\epsilon)\rho_{\rm CTC}^i = \epsilon \rho \geq 0$, $(1+\epsilon)\rho_{\rm CTC}^i-\rho_{\rm CTC}^f = \epsilon (2\rho_{\rm CTC}^i-\rho)$ 
is non-negative when $2\rho_{\rm CTC}^i-\rho \geq 0$. Thus, $\| \rho_{\rm CTC}^f-\rho_{\rm CTC}^i \| \leq \epsilon$ holds when 
$2\rho_{\rm CTC}^i \geq \rho$. We refer to this relation as the ``approximate teleportation condition''.
In the special case that we are working in, which is when Eq. (\ref{eq:final-state}) is valid, the approximate teleportation condition implies that $\rho_{\rm CTC}^i=\frac{I_2}{2}$, since \(\rho\) is an arbitrary state, and where we have additionally assumed that the input state to be teleported is a qubit. 
Here \(I_2\) is the identity operator on the qubit Hilbert space. 
Let $\rho=|\psi\rangle \langle \psi|$ be an arbitrary single-qubit pure state, with 
$|\psi\rangle = \cos\frac{\theta}{2}|0\rangle + e^{i\phi}\sin\frac{\theta}{2}|1\rangle$, on the Bloch sphere, so that the average fidelity is
\begin{equation}
\int \langle \psi|\rho_{\rm CTC}^f|\psi\rangle d\psi=\frac12(1+\epsilon),
\end{equation}
which exceeds the classical fidelity of two-thirds, when $\epsilon > \frac13$.\\

\noindent We note here that when \(\epsilon = 0\), the combined CR-CTC system is not quantum mechanical. It may however be expected that for nonzero \(\epsilon\), quantum mechanics will be regained. Our results show that quantum mechanics is at least beyond \(\epsilon = \frac{1}{3}\).\\

\section{Conclusion}
\label{sec-conclu}

\noindent In summary, we find that cloning and deleting of quantum information are possible in causality respecting quantum systems provided we allow it to interact with a 
closed timelike curve. These phenomena are impossible if we disallow the interaction with CTC. 
We show that the phenomena of cloning and deleting are possible  in the presence of CTCs  due to two other phenomena, 
viz.,
popping up and elimination of quantum information, that seem to be natural to CR systems in the presence of CTCs. We moreover prove that teleportation of 
an arbitrary quantum state, 
even approximately, from a CR system to a CTC cannot occur, although teleportation between two CR systems is known to be possible. 
We also find limits to the creation of entangled states between a CR system and a CTC, and between two CRs in the presence of a CTC.\\

\noindent It is interesting to mention here a result by Hardy \cite{Hardy}, which shows that it is possible to build a toy model of the physical world
in which teleportation is possible and 
no-cloning is valid, while the system has a local realistic description. It was later shown \cite{noncommuting} 
that if one considers an arbitrary set of commuting quantum states, then any 
member of the set can be teleported through a separable shared quantum state. 
The shared quantum state, being separable, has a local hidden variable description, and an arbitrary set of 
commuting quantum states, since it may contain nonorthogonal quantum states, that cannot be cloned (in causality respecting quantum systems), 
although can be (quantum mechanically) broadcast \cite{broadcasting}. In this paper, we found that a CR system in the presence of a CTC 
has the opposite behavior: Teleportation is disallowed, while cloning is possible.\\

\noindent We subsequently show that a modification in the Deutsch model of CTCs, which we call the \(\epsilon\)-close Deutsch model, allows for nonclassical teleportation fidelities. We believe that our results show the power as well as the limitations of CTC in quantum information processing.

\begin{acknowledgments}
We acknowledge discussions with Sasha Sami. 
\end{acknowledgments}


\begin{thebibliography}{99}
\bibitem{imp-note} With regard to the applicability of the Deutsch criterion, we hold the
following belief. The paper of Deutsch \cite{deu} does not rule out this
belief. Rather to the contrary, it seems that this is a natural way to use
the Deutsch-CTC evolution. The Deutsch formalism of closed time-like
curves does not define the ``point-of-exit'' of a time traveler from such
a curve. Therefore, the Deutsch self-consistency condition should be
applicable also for intermediate times during the evolution.

A typical scenario can be the following. An observer, Ayesha, applies the
joint unitary \(U_A\) on the joint CR-CTC system. If she wishes to know
the output, she must apply the Deutsch consistency equation. [Indeed, the
output should not depend on whether or not she ``wishes'' to know the
output.] It is not yet known whether another observer will have her/his
hands on the joint system. However, it so happens that another observer,
Badal, gets hold of the joint systems, and promptly applies the joint
unitary \(U_B\) on the joint CR-CTC system. If he too wishes to know the
output, he needs to apply the Deutsch equation again.

There are certainly other options to proceed. In one, Ayesha needs to
ascertain that no other observer is going to apply a joint unitary on the
CR-CTC system before she considers the Deutsch condition. In another,
after Badal gets hold of the joint system, he needs to know whether Ayesha
had applied the Deutsch condition beforehand, and whether later observers
(after Badal) will apply any joint unitary. We believe that this situation
is too artificial for any evolution.

It is difficult to ascertain when ``all'' interactions have been
performed, and why a future observer should agree to such a restriction.
Even if we decide today that all operations are complete (“we” being the
experimenter), and if we are allowed to take this decision, we can decide
tomorrow that it is not. If this change of decision is not allowed, that
would be an extra restriction on the Deutsch evolution rules.

We have therefore taken the avenue that intermediate application of the
Deutsch criterion is allowed.

Please also look at the Remarks 1.1, 2.1 and 2.2 made in the main text,
where we clearly state the exact form in which the Deutsch condition is
applied in the corresponding proofs.
\end{thebibliography}
\end{document}